\documentclass[prd,preprint,showpacs,showkeys,footinbib]{revtex4}
\usepackage{amsmath}
\usepackage{graphics}
\usepackage{isotope}

\renewcommand\vec[1]{\mathbf{#1}}
\newcommand\he[1]{#1^{\dagger}}
\DeclareMathOperator{\re}{Re}
\DeclareMathOperator{\im}{Im}
\DeclareMathOperator{\tr}{Tr}
\newcommand\bra[1]{\langle#1|}
\newcommand\ket[1]{|#1\rangle}
\newcommand\vev[1]{\langle#1\rangle}
\newcommand\doublet[2]{\genfrac{\{}{\}}{0pt}{}{#1}{#2}}

\begin{document}

\title{Goldstone bosons in presence of charge density}
\author{Tom\'{a}\v{s} Brauner}
\email{brauner@ujf.cas.cz}
\affiliation{Department of Theoretical Physics,
Nuclear Physics Institute ASCR, 25068 \v Re\v z, Czech Republic}

€\begin{abstract} We investigate spontaneous symmetry breaking in
Lorentz-noninvariant theories. Our general discussion includes relativistic
systems at finite density as well as intrinsically nonrelativistic systems. The
main result of the paper is a direct proof that nonzero density of a
non-Abelian charge in the ground state implies the existence of a Goldstone
boson with nonlinear (typically, quadratic) dispersion law. We show that the
Goldstone boson dispersion relation may in general be extracted from the
current transition amplitude and demonstrate on examples from recent
literature, how the calculation of the dispersion relation is utilized by this
method. After then, we use the general results to analyze the nonrelativistic
degenerate Fermi gas of four fermion species. Due to its internal
$\mathrm{SU(4)}$ symmetry, this system provides an analog to relativistic
two-color quantum chromodynamics with two quark flavors. In the end, we extend
our results to pseudo-Goldstone bosons of an explicitly broken symmetry.
\end{abstract}

\maketitle

\section{Introduction}
The basic result concerning the physics of spontaneous breaking of global
internal continuous symmetries in Lorentz-noninvariant systems was achieved by
Nielsen and Chadha thirty years ago \cite{Nielsen:1976hm}. They showed that,
under certain technical assumptions (the most important one being that of a
sufficiently fast decrease of correlations, which ensures that no long-range
forces are present), the energy of the Goldstone boson (GB) of the
spontaneously broken symmetry is proportional to some power of momentum in the
low-momentum limit, and classified the GBs as type-I and type-II according to
whether this power is odd or even, respectively. Their general counting rule
states that \emph{the number of type-I GBs plus twice the number of type-II GBs
is greater than or equal to the number of broken generators.}

Two classic examples of systems where the type-II GBs occur are the ferromagnet
and the so-called $A$ phase of superfluid \isotope[3]{He}. The issue of GBs in
Lorentz-noninvariant theories has regained considerable interest in the past
decade due to the discovery of several other systems exhibiting the existence
of type-II GBs. These fall into two wide classes: Relativistic systems at
finite density and intrinsically nonrelativistic systems. The first class
includes various color-superconducting phases of dense quark matter---the
color-flavor-locked (CFL) phase with a kaon condensate
\cite{Miransky:2001tw,Schaefer:2001bq} and the two-flavor color superconductor
\cite{Blaschke:2004cs}---as well as the Bose--Einstein condensation of
relativistic Bose gases \cite{Andersen:2005yk,Andersen:2006ys} and the
relativistic nuclear ferromagnet \cite{Beraudo:2004zr}. The second class covers
the before mentioned ferromagnet and superfluid helium, and the rapidly
developing field of the Bose--Einstein condensation of dilute atomic gases with
several internal degrees of freedom \cite{He:2006ne}.

Besides the investigation of particular systems, a few new general results also
appeared. Leutwyler \cite{Leutwyler:1994gf} showed within the low-energy
effective field theory framework that the nonlinearity of the GB dispersion is
tightly connected to the fact that some of the conserved charges develop
nonzero density in the ground state. As a matter of fact, the nonzero charge
density induces breaking of time-reversal invariance, which in turn leads to
the appearance of a term in the effective Lagrangian with a single time
derivative. This term is then responsible for the modification of the GB
dispersion. In addition to Leutwyler's results, Schaefer \emph{et al.}
\cite{Schaefer:2001bq} proved that nonzero ground-state density of a commutator
of two broken charges is a necessary condition for an abnormal number of GBs.

In our previous work \cite{Brauner:2005di} we showed that, at least within a
particular class of systems described by the relativistic linear sigma model at
finite density, and at tree level, the converse to the theorem of Schaefer
\emph{et al.} is also true: Nonzero density of a commutator of two broken
charges implies the existence of a single GB which has quadratic dispersion
relation, i.e., is type-II. Moreover, there is a one-to-one correspondence
between the pairs of broken generators whose commutator has nonzero density,
and the type-II GBs. In Ref. \cite{Brauner:2006xm} we analyzed the particular
model of Miransky and Shovkovy \cite{Miransky:2001tw} and Schaefer \emph{et
al.} \cite{Schaefer:2001bq} at one-loop level and asserted that the one-loop
corrections do not alter the qualitative conclusions achieved at the tree
level.

The purpose of this paper is to extend the previous results in a
model-independent way. We are going to present a general argument, based on a
Ward identity for the broken symmetry, that nonzero ground-state density of a
non-Abelian charge leads to the existence of a GB with nonlinear (typically,
quadratic) dispersion law. The plan of the paper is following. In the next
section we give a detailed derivation of the main result. In the subsequent
sections we then apply it to several systems discussed in recent literature,
and show how it facilitates the calculation of the GB dispersion relations.
Next we analyze in detail Cooper pairing in a nonrelativistic gas of four
degenerate fermion species. Finally, we show how the proposed method may be
generalized to include explicit symmetry breaking.

\section{General argument}
\label{Sec:general} In Ref. \cite{Brauner:2005di} we provided a general
argument that nonzero density of a non-Abelian charge leads to modified GB
counting. In particular, we invoked the standard proof of the Goldstone theorem
\cite{Goldstone:1962es} to show that
\begin{equation}
if_{abc}\bra0j^0_c(x)\ket0=\bra0[j^0_a(x),Q_b]\ket0
=2i\im\sum_n\bra0j^0_a(0)\ket n\bra nj^0_b(0)\ket0,
\label{charge_order_parameter}
\end{equation}
see Eq. (1) in Ref. \cite{Brauner:2005di} and the discussion below it. Here
$f_{abc}$ is the set of structure constants of the symmetry group and the index
$n$ runs over different zero-momentum states in the spectrum. The vacuum
expectation value $\bra0j^0_c\ket0$ represents the order parameter for symmetry
breaking, while both currents $j^0_a$ and $j^0_b$ serve as interpolating fields
for the GBs. $Q_b$ is the charge operator associated with the current
$j^{\mu}_b$. Note that whereas in Lorentz-invariant systems there is a
one-to-one correspondence between the GBs and the broken currents, at nonzero
charge density the situation changes. As Eq. \eqref{charge_order_parameter}
clearly exhibits, a single GB now couples to the two broken currents whose
commutator has nonzero vacuum expectation value. Before we expand this simple
observation into a more detailed argument, we discuss what the involved current
transition amplitudes tell us about the GB dispersion relation.

\subsection{GB dispersion from transition amplitude}
The matrix element which couples the GB state $\ket{\pi(\vec k)}$ to the broken
current operator $j^{\mu}_a(0)$, may be generally parameterized as
\begin{equation}
\bra0j^{\mu}_a\ket{\pi(\vec k)}=ik^{\mu}F_a(|\vec k|)+i\delta^{\mu0}G_a(|\vec
k|).
\label{current_amplitude_parametrization}
\end{equation}
By the second term, $G_a$, we explicitly account for the possibility of Lorentz
violation. However, we still assume that the rotational invariance remains
intact. The current conservation then implies the following equation for the GB
dispersion (see also Ref. \cite{Hofmann:1998pp}), $\omega(\vec k)$,
\begin{equation}
(\omega^2-\vec k^2)F_a+\omega G_a=0.
\label{general_dispersion}
\end{equation}

Let us now explicitly mention a few special cases, distinguished by the
limiting behavior of the functions $F_a$ and $G_a$ at low momentum. First, when
$G_a=0$ for all values of $\vec k$, we recover the standard Lorentz-covariant
parameterization of the transition amplitude
\eqref{current_amplitude_parametrization}, and the standard Lorentz-invariant
dispersion relation, $\omega=|\vec k|$. Second, suppose that $G_a/F_a$ is small
in the limit $|\vec k|\to0$, of order $\mathcal O(|\vec k|)$. Then the
dispersion relation comes out linear, only the phase velocity is no more equal
to one due to the Lorentz-violating term $G_a$. Third and most importantly for
the rest of the paper, when $G_a$ has a finite limit, $\lim_{|\vec
k|\to0}G_a\neq0$, the dispersion relation takes the low-momentum form
\begin{equation}
\omega(\vec k)=\vec k^2\frac{F_a}{G_a}.
\label{quadratic_dispersion}
\end{equation}
Eq. \eqref{quadratic_dispersion} implies that the GB dispersion relation is
nonlinear. However, in order to determine the corresponding power of momentum
and thus classify the GB according to the Nielsen--Chadha scheme, we would have
to further know the limiting behavior of $F_a$.

One should keep in mind that so far, we have provided just a parameterization of
the GB dispersion relation in terms of the current transition amplitude, which
only assumed rotational invariance and current conservation. In order to decide
which of the possibilities outlined above is actually realized, we need more
information on the dynamics of the system, i.e., on the functions $F_a,G_a$.

\subsection{Transition amplitude from Ward identity}
We now get back to Eq. \eqref{charge_order_parameter}. In order to turn it into
a quantitative prediction of the GB dispersion relation, we replace the
commutator with the time ordering, that is, consider the correlator of two
currents, $\bra0T\{j_a^{\mu}(x)j_b^{\nu}(y)\}\ket0$. Taking a divergence and
using the current conservation, we arrive at a simple Ward identity,
\begin{equation}
\partial^x_{\mu}\bra0T\{j_a^{\mu}(x)j_b^{\nu}(y)\}\ket0=\delta^4(x-y)if_{abc}
\bra0j^{\nu}_c(x)\ket0.
\label{Ward_identity}
\end{equation}
Of course, with the assumed rotational invariance the right hand side of Eq.
\eqref{Ward_identity} may only be nonzero for $\nu=0$ so that it exactly
encompasses our order parameter---the charge density.

Throughout the paper we assume that the symmetry is not anomalous so that, in
particular, the naive Ward identity \eqref{Ward_identity} holds without any
corrections due to, e.g., Schwinger terms. This assumption is partially
justified by the models studied in the following sections, where our results
prove equivalent to those obtained with different methods. In general, however,
the validity of Eq. \eqref{Ward_identity} should be checked case by case.

Next we use the K\"all\'en--Lehmann representation for the current--current
correlator in the momentum space, plugging in the parameterization
\eqref{current_amplitude_parametrization}. We find that the condition
\eqref{general_dispersion}, as expected, ensures cancelation of the
one-particle poles in the correlator. Taking the limit $|\vec k|\to0$, it
follows that in order to get a finite nonzero limit as the right hand side of
the Ward identity \eqref{Ward_identity} requires, the functions $G_a,G_b$ have
to have a nonzero limit as well. The residua of the poles then yield the
density rule $G_aG^*_b-G^*_aG_b=if_{abc}\bra0j^0_c\ket0$, or,
\begin{equation}
\im G_aG_b^*=\frac12f_{abc}\bra0j^0_c\ket0,
\label{density_rule}
\end{equation}
where the $G$s now stand for the zero-momentum limits of the respective
functions defined by Eq. \eqref{current_amplitude_parametrization}. Eq.
\eqref{density_rule} together with Eq. \eqref{quadratic_dispersion} constitute
the main result of the paper. They say that nonzero density of a non-Abelian
charge in the ground state requires the existence of one GB with nonlinear
dispersion relation. In general (in fact, in all cases the author is aware of),
this dispersion is quadratic unless $F_a$ vanishes at $\vec k=0$. We will
speculate on this possibility in the conclusions.

A few other comments are in order here. First, the fact that the function $G_a$
has nonzero zero-momentum limit essentially means that the broken charge
operator creates the GB even at zero momentum with a finite probability. This
is in contrast to the Lorentz-invariant case. Second, in dense relativistic
systems the function $G_a$ must be proportional to the only Lorentz-violating
parameter in the theory, the chemical potential. An example will be given in
the next section. Last, in view of Eqs. \eqref{quadratic_dispersion} and
\eqref{density_rule}, it is tempting to conclude that the energy of the GB is
inversely proportional to the charge density. This indeed happens in the
ferromagnet \cite{Leutwyler:1994gf}, but need not be true generally. Again in
the next section, we shall see that both $F_a$ and $G_a$ may be proportional to
the charge density so that it drops in the ratio.

\section{Linear sigma model}
\label{Sec:linear_sigma_model} The first class of systems to which we shall now
apply our general results will be that described by the relativistic linear
sigma model at finite density (or chemical potential). This was already studied
in Ref. \cite{Brauner:2005di} where we found that the chemical potential
induces mixing terms in the Lagrangian and the excitation spectrum can be
determined only upon an appropriate diagonalization of the matrix propagator.
In this section we shall see that this complication may be circumvented by the
use of Eq. \eqref{general_dispersion}, at least in the case of type-II GBs. We
start with a detailed analysis of a simple example in order to illustrate the
validity of the formulas \eqref{quadratic_dispersion} and \eqref{density_rule}.

\subsection{Model with $\mathrm{SU(2)\times U(1)}$ symmetry}
\label{Sec:Miransky_model}
We recall the model that was used to describe kaon condensation in the CFL
phase of dense quark matter \cite{Miransky:2001tw,Schaefer:2001bq}. It is
defined by the Lagrangian,
\begin{equation}
\mathcal L=\he{D_{\mu}\phi}D^{\mu}\phi-M^2\he\phi\phi-\lambda(\he\phi\phi)^2,
\label{Lagrangian_Shovkovy}
\end{equation}
where the scalar field $\phi$ transforms as a complex doublet of the global
$\mathrm{SU(2)}$ symmetry. In order to account for finite density, the chemical
potential $\mu$ associated with the global $\mathrm{U(1)}$ invariance (particle
number) is introduced via the covariant derivative,
$D_0\phi=(\partial_0-i\mu)\phi$.

Once the value of the chemical potential exceeds the mass parameter $M$, the
scalar field condenses. The ground state may be chosen so that only the lower
component of $\phi$ develops expectation value. At tree level,
$\vev{\phi_2}=v/\sqrt2$, where $v=\sqrt{(\mu^2-M^2)/\lambda}$. This breaks the
original $\mathrm{SU(2)\times U(1)}$ symmetry to its $\mathrm{U(1)_Q}$
subgroup, generated by the matrix $Q=\frac12(\openone+\tau_3)$. The spectrum
contains two (as opposed to three broken generators) GBs. The type-II GB is
annihilated by $\phi_1$ and its exact (tree-level) dispersion relation is given
by
\begin{equation}
\omega=\sqrt{\vec k^2+\mu^2}-\mu.
\label{typeIIdispersion}
\end{equation}
The type-I GB is annihilated by a (energy-dependent) linear combination of
$\phi_2$ and $\he\phi_2$, and its dispersion relation corresponds to the
``$-$'' case of
\begin{equation}
\omega^2_\pm(\vec k)=\vec k^2+3\mu^2-M^2\pm\sqrt{4\mu^2\vec
k^2+(3\mu^2-M^2)^2}. \label{typeIdispersion}
\end{equation}

After shifting the field $\phi_2$ by its vacuum expectation value, the Noether
currents $j_a^{\mu}$ associated with the $\mathrm{SU(2)}$ symmetry of the model
pick up terms linear in $\phi$. From these, we may readily express the current
transition amplitudes related to the type-II GB state $\ket{G(\vec k)}$ as
\begin{equation}
\bra0j^{\mu}_1\ket{G(\vec k)}=-i\bra0j^{\mu}_2\ket{G(\vec k)}=
\frac v{\sqrt2}(k^{\mu}+2\mu\delta^{\mu0})\bra0\phi_1(0)\ket{G(\vec k)}.
\label{typeIIcurrent_amplitudes}
\end{equation}
Note that this, according to Eq. \eqref{quadratic_dispersion}, immediately
leads to the dispersion relation $\omega=\vec k^2/2\mu$, which is indeed the
low-momentum limit of the exact result \eqref{typeIIdispersion}. Moreover,
evaluating the amplitude $\bra0\phi_1(0)\ket{G(\vec k)}$ explicitly gives
$$
G_1G_2^*=-2i\mu^2v^2|\bra0\phi_1(0)\ket{G(\vec k)}|^2=-i\mu v^2
$$
at $|\vec k|=0$. This is in accord with the density rule \eqref{density_rule}
for the ground-state isospin density is simply given by the constant term in
$j_3^0$, $\bra0j^0_3\ket0=-\mu v^2$, and the relevant structure constant of the
$\mathrm{SU(2)}$ group in the basis of Pauli matrices is $f_{123}=2$.

Analogously, the coupling of the type-I GB state $\ket{\pi(\vec k)}$ to the
current $j^{\mu}_3$ is found to be
\begin{equation*}
\bra0j^{\mu}_3\ket{\pi(\vec k)}=\frac
v{\sqrt2}\left[(k^{\mu}-2\mu\delta^{\mu0})
\bra0\he\phi_2(0)\ket{\pi(\vec k)}
-(k^{\mu}+2\mu\delta^{\mu0})
\bra0\phi_2(0)\ket{\pi(\vec k)}\right].
\end{equation*}
The basic complication as opposed to the type-II GB is that now the current
amplitude is not proportional to a single scalar field amplitude as in Eq.
\eqref{typeIIcurrent_amplitudes}, but contains two entangled amplitudes.
Therefore, these do not simply drop in the ratio as before and have to be
evaluated explicitly. Using the expressions for the scalar field amplitudes
derived in Ref. \cite{Brauner:2006xm}, the final result for the transition
amplitudes at $|\vec k|\to0$ becomes
\begin{equation*}
\begin{split}
\bra0j^0_3\ket{\pi(\vec k)}&=-\frac v{\sqrt2}\sqrt{|\vec k|}
\left(\frac{\mu^2-M^2}{3\mu^2-M^2}\right)^{-1/4},\\
\bra0\vec j_3\ket{\pi(\vec k)}&=-\frac v{\sqrt2}\frac{\vec k}{\sqrt{|\vec
k|}}\left(\frac{\mu^2-M^2}{3\mu^2-M^2}\right)^{1/4}.\\
\end{split}
\end{equation*}
A simple ratio of the coefficients of the spatial and temporal current
transition amplitudes gives the dispersion relation
$$
\omega=\sqrt{\frac{\mu^2-M^2}{3\mu^2-M^2}}|\vec k|,
$$
which is the correct low-momentum limit of the full dispersion
\eqref{typeIdispersion}.

\subsection{General case}
We checked the formulas \eqref{general_dispersion} and \eqref{density_rule} on
the simple example of the linear sigma model with an $\mathrm{SU(2)\times
U(1)}$ symmetry. We saw that their application is particularly simple for the
type-II GBs, which carry unbroken charge. On the other hand, the application to
type-I GBs is obscured by mixing of scalar field transition amplitudes.

With this experience we are ready to investigate the general linear sigma model
with arbitrary symmetry. In Ref. \cite{Brauner:2005di}, we dealt with the
Lagrangian
$$
\mathcal L=D_{\mu}\he\phi D^{\mu}\phi-V(\phi),
$$
where $V(\phi)$ is the most general static potential containing terms up to
fourth order in $\phi$ that is invariant under the prescribed symmetry. The
covariant derivative, $D_{\mu}\phi=(\partial_{\mu}-iA_{\mu})\phi$, includes the
chemical potential(s) assigned to one or more mutually commuting generators of
the symmetry group.

Once the chemical potential is large enough, the scalar field develops nonzero
expectation value, $\phi_0$, and must be reparameterized as
\begin{equation}
\phi(x)=e^{i\Pi(x)}[\phi_0+H(x)].
\label{general_reparameterization}
\end{equation}
The field $\Pi(x)$ is a linear combination of the broken generators and
includes the Goldstone degrees of freedom. The ``radial'' degrees of freedom
are described by $H(x)$.

The Noether current associated with a particular conserved charge $T_a$ reads
$$
j_a^{\mu}=-2\im\he\phi T_aD^{\mu}\phi=-2\im\he\phi
T_a\partial^{\mu}\phi+2\he\phi T_aA^{\mu}\phi.
$$
Now we just have to insert the parameterization
\eqref{general_reparameterization} and retain the terms linear in the field
$\Pi$,
$$
j^{\mu}_{a,\Pi,\text{lin}}=
-\he\phi_0\{T_a,\partial^{\mu}\Pi\}\phi_0+2i\he\phi_0A^{\mu}[T_a,\Pi]\phi_0.
$$
According to Eq. \eqref{quadratic_dispersion}, this is all we need to calculate
the dispersions of the type-II GBs present in the system. The commutator and
anticommutator will give factors that are purely geometrical, i.e., determined
by the structure of the symmetry group and the symmetry breaking pattern.
Without having to evaluate them explicitly, we simply observe that the constant
($G_a$) term is proportional to the chemical potential. This means that the
dispersion relation always comes out as $\omega=c\vec k^2/\mu$, where $c$ is
some, yet unknown $\mu$-independent constant.

This constant can be determined in a different manner. Recall that (at least at
tree level), once there is a (at least discrete) symmetry that prevents the
potential $V(\phi)$ from picking a cubic term, the phase transitions between
the normal and the Bose--Einstein condensed phases as well as those between two
ordered phases are of second order. This means that the dispersion relations of
the excitations must be continuous across the phase transitions. Hence, due to
their particularly simple form, they can be matched to the dispersion in the
normal phase. In the normal phase, a particle carrying charge $Q$ of the
symmetry equipped with the chemical potential $\mu$, has dispersion
$\omega=\sqrt{\vec k^2+M^2}-\mu Q$, $M$ being its gap at $\mu=0$. At $M=\mu Q$
the particle becomes a GB of the spontaneously broken symmetry and its
low-momentum dispersion relation reads
\begin{equation}
\omega=\frac{\vec k^2}{2\mu Q}.
\label{typeII_dispersion_Q}
\end{equation}
Due to the argument sketched above, this dispersion relation then persists to
the broken-symmetry phase for arbitrary values of the chemical potential.

\section{Nambu--Jona-Lasinio model}
\label{Sec:NJL_model} Now we turn our attention to models of the
Nambu--Jona-Lasinio (NJL) type, i.e., purely fermionic models with local,
nonderivative four-fermion interaction. There, symmetry is broken dynamically
by a vacuum expectation value of a composite operator. At the same time, the GB
of the spontaneously broken symmetry is a bound state of the elementary
fermions, and it manifests itself as a pole in the correlation functions of the
appropriate composite operators.

The standard way to calculate the dispersion relation of the GB is as follows
(see e.g. Ref. \cite{Blaschke:2004cs}). One first performs the
Hubbard--Stratonovich transformation, including the corresponding pairing
channels. The purely scalar action is next expanded to second order in the
auxiliary scalar field so as to get the GB propagator and find its pole or,
equivalently, the zero of its inverse. However, the inverse propagator contains
a constant term which prevents us from identifying the gapless pole directly.
This term is removed with the help of the gap equation. Only after then may the
inverse propagator be expanded to second power in momentum in order to
establish the dispersion relation.

The approach proposed in this paper simplifies the calculation of the GB
dispersion relation in three aspects. First, there is no need to use the gap
equation in order to establish the existence of a gapless excitation. Indeed,
since the only assumption, besides rotational invariance, is the current
conservation, the broken symmetry is already built into the formalism. Second,
the loop integrals are only expanded to first order in the momentum, again
since one power of momentum is already brought about by the current
conservation. This saves us one order in the Taylor expansion and thus reduces
the analytic formulas considerably. (Of course, the final result must be the
same whatever method we use. We just choose the less elaborate one.) Third, at
least when calculating the dispersion relation of a type-II GB, we may take the
zero-th component of the external momentum equal to zero from the very
beginning. This means that, at finite temperature, we get a direct expression
of the GB dispersion in terms of thermal loops, without having to analytically
continue the result back to Minkowski spacetime afterwards.

\subsection{Dispersion relation from fermion loops}
\label{Sec:fermion_loops} In the models of the NJL type, the Noether currents
are fermion bilinears. Their couplings to the Goldstone states are therefore
given by the one-loop diagrams, symbolically,
$$
\bra0j^{\mu}\ket{\pi(\vec
k)}=\scalebox{0.8}{\parbox{20mm}{\includegraphics{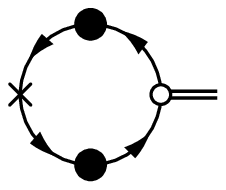}}}
$$
The solid circles denote the full fermion propagators, which are in turn
obtained by solving the gap equation, while the empty circle stands for the yet
unknown vertex coupling the two fermions to the GB state. In case of a type-II
GB the dispersion relation is then, according to Eqs.
\eqref{current_amplitude_parametrization} and \eqref{quadratic_dispersion},
given by the ratio,
\begin{equation}
\omega(\vec k)=\frac{\vec k\cdot\vec
j_a\scalebox{0.8}{\parbox{20mm}{\includegraphics{fig1}}}}
{j^0_a\scalebox{0.8}{\parbox{20mm}{\includegraphics{fig1}}}},
\label{loop_ratio}
\end{equation}
where the symbols standing in front of the loops denote the operators inserted
in place of the crosses.

Before we explain in detail how the Goldstone--fermion vertex may be extracted
from another Ward identity, let us note that the explicit evaluation of this
vertex may in fact often be circumvented. This will be demonstrated later on
explicit examples. In general, the auxiliary scalar fields introduced by the
Hubbard--Stratonovich transformation mix so that one has to look for poles of a
matrix propagator. However, it seems to be a generic feature of type-II GBs
that they carry charge of some unbroken symmetry. As a result, they may couple
to just one of the auxiliary scalar fields and no mixing appears. (For
instance, recall the model of Sec. \ref{Sec:Miransky_model} where the type-II
GB couples just to $\phi_1$, while the type-I GB couples to both $\phi_2$ and
$\he\phi_2$.) In other words, we may devise an interpolating field $\he\psi
X\psi$, $X$ being some matrix in the flavor space, and calculate instead of the
transition amplitude $\bra0j^{\mu}\ket{\pi(\vec k)}$ the correlation function
of $j^{\mu}$ and $\he\psi X\psi$. The point is that, by the K\"all\'en--Lehmann
representation, this correlation function is proportional to the desired
transition amplitude, and that the GB propagator as well as the unknown
amplitude $\bra{\pi(\vec k)}\he\psi X\psi\ket0$ drop out in the ratio
\eqref{loop_ratio}.

In practice this means that, instead of the unknown Goldstone--fermion vertex
in the loops in Eq. \eqref{loop_ratio}, we may simply insert the operator
$\he\psi X\psi$. In particular, in the nonrelativistic NJL model, the current
$j^{\mu}_a$ associated to the charge $T_a$ reads
$$
j^0_a=\he\psi T_a\psi,\quad
\vec j_a=\frac1{2im}(\he\psi T_a\nabla\psi-\nabla\he\psi T_a\psi),
$$
where $m$ is the common mass of the degenerate fermion flavors. Inserting these
operators into the fermion loops in Eq. \eqref{loop_ratio}, and making the
Taylor expansion of the spatial loop to the first order in the external
momentum, we arrive at the result
\begin{equation}
\omega(\vec k)=\frac{\vec k^2}{2m}\Biggl\{
1+\frac43\frac{\int\frac{d^4q}{(2\pi)^4}\tr\bigl[\vec q^2\frac{\partial S}
{\partial\vec q^2}(q)XS(q)T_a\bigr]}
{\int\frac{d^4q}{(2\pi)^4}\tr\left[S(q)XS(q)T_a\right]}\Biggr\},
\label{NR_dispersion_formula}
\end{equation}
where the trace is performed over the flavor space and $S$ denotes the full
fermion propagator. At finite temperature $1/\beta$ the integral over $q^0$ in
Eq. \eqref{NR_dispersion_formula} becomes a sum over fermion Matsubara
frequencies, $\nu_n=(2n+1)\pi/\beta$. The GB thus naturally remains gapless up
to a possible thermal symmetry restoring phase transition, but otherwise its
dispersion can depend on temperature.

\subsection{GB--fermion coupling from Ward identity}
As already stressed in the discussion below Eq. \eqref{loop_ratio}, the formula
\eqref{NR_dispersion_formula} has a limited use. It applies only to the cases
in which the GB couples to just one of the auxiliary scalar fields introduced
by the Hubbard--Stratonovich transformation (no mixing). Unless this happens,
the (unknown) couplings of the GB to the fermion bilinear operators do not drop
in the ratio \eqref{loop_ratio}.

We also noted that this no-mixing property is most likely to be generic for
type-II GBs, which carry charge of some unbroken symmetry. Nevertheless, in any
case, the GB--fermion coupling can be evaluated directly by using yet another
Ward identity \cite{Brauner:2005hw}. Let us introduce the connected vertex
function, $G^{\mu}_a(x,y,z)=\bra0T\{j^{\mu}_a(x)\psi(y)\he\psi(z)\}\ket0$.
Taking the divergence and stripping off the full fermion propagators
corresponding to the external legs, we find the Ward identity for the proper
vertex,
\begin{equation}
k_{\mu}\Gamma_a^{\mu}(p+k,p)=T_aS^{-1}(p+k)-S^{-1}(p)T_a.
\label{WI_Margolis}
\end{equation}

This vertex exhibits a pole due to the propagation of the intermediate GB
state. Near the GB mass shell, it may be approximated by a sum of the bare
vertex and the pole contribution \cite{Jackiw:1973tr}. Writing the inverse
nonrelativistic fermion propagator as $S^{-1}(\omega,\vec p)=\omega-\frac{\vec
p^2}{2m}+\mu-\Sigma(\omega,\vec p)$, where $\Sigma$ is the self-energy, the
bare part of the propagator exactly cancels against the bare vertex in the Ward
identity \eqref{WI_Margolis}. The remaining term expresses the residuum at the
GB pole in the vertex function---which is proportional to the desired
GB--fermion coupling---in terms of the fermion self-energy, so that
\begin{equation}
\scalebox{0.8}{\parbox{15mm}{\includegraphics{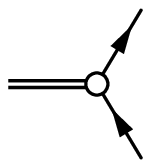}}}=\frac1N
\left[\Sigma(p)T_a-T_a\Sigma(p+k)\right].
\label{GB_fermion_coupling}
\end{equation}
The normalization constant $N$ is proportional to the coupling of the GB to the
broken current itself, but this is not important since this factor only appears
in the ratio \eqref{loop_ratio}. The GB--fermion coupling is apparently
determined by the symmetry-breaking part of the fermion self-energy. Once this
is calculated, e.g., in the mean-field approximation, it may be plugged into
the loops in Eq. \eqref{loop_ratio} to determine the GB dispersion.

In conclusion of our general discussion let us remark that all the results
presented so far in Sec. \ref{Sec:NJL_model} extend straightforwardly to cases
including Cooper pairing of fermions. Once the GB is annihilated by a difermion
operator instead of a fermion--antifermion one, we just have to flip the
direction of some of the arrows in the fermion loops. The $S(q)$ in Eq.
\eqref{NR_dispersion_formula} then denotes the matrix propagator in both the
flavor and the Nambu ($\psi,\he\psi$) space.

In addition, Eqs. \eqref{WI_Margolis} and \eqref{GB_fermion_coupling} modify
accordingly: The matrix $T_a$ must be replaced with the block matrix in the
Nambu space, $\left(\begin{smallmatrix}T_a & 0\\0 &
-T_a^T\end{smallmatrix}\right)$. In the end, the GB--fermion coupling vertex
comes out as
\begin{equation}
\scalebox{0.8}{\parbox{15mm}{\includegraphics{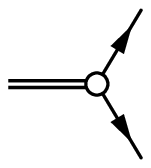}}}=-\frac1N
\left[\Sigma_{\psi\psi}(p)T_a^T+T_a\Sigma_{\psi\psi}(p+k)\right],
\label{GB_fermion_coupling_Cooper}
\end{equation}
where $\Sigma_{\psi\psi}$ denotes the anomalous (pairing) part of the fermion
self-energy.

\subsection{Three-flavor degenerate Fermi gas}
The nonrelativistic Fermi gas consisting of three degenerate fermion species
has been investigated theoretically in great detail
\cite{Honerkamp:2003hh,Honerkamp:2004ho,He:2006ne}. The reason is that such a
system has already been successfully realized in experiments using atoms of
\isotope[6]{Li} and \isotope[40]{K}. At the same time it provides an
interesting analogy with quantum chromodynamics (QCD). A recent theoretical
analysis suggests that under certain conditions, bound clusters of three atoms
may form, being singlets with respect to the global $\mathrm{SU(3)}$ symmetry,
very much like baryons in QCD \cite{Rapp:2006rx}. Such a three-flavor atomic
Fermi gas thus creates a natural playground for testing properties of quark
matter.

In the following, we are going to illustrate the utility of our general
formulas and rederive the results of Ref. \cite{He:2006ne} concerning the
type-II GBs in the three-flavor system. To that end, recall that with a proper
choice of the interaction, the global $\mathrm{SU(3)}$ symmetry is broken by a
difermion condensate, $\langle\psi_i\psi_j\rangle$. Since this is antisymmetric
under the exchange of $i,j$, it transforms in the $\overline{\mathbf 3}$
representation of $\mathrm{SU(3)}$. This leads to the symmetry breaking
pattern $\mathrm{SU(3)\times U(1)\to SU(2)\times U(1)}$.

Once we choose the condensate to point in the third direction, the generator
$\lambda_8$ acquires nonzero vacuum expectation value. Based on the general
discussion in Sec. \ref{Sec:general}, we anticipate that the four broken
generators, $\lambda_4,\lambda_5,\lambda_6,\lambda_7$, couple to a doublet of
type-II GBs with quadratic dispersion relations. Due to the unbroken symmetry,
these two GBs have the same dispersion. For the explicit calculation, we choose
the GB in the sector $(\lambda_4,\lambda_5)$.

\begin{figure}
\begin{center}
\scalebox{1}{\includegraphics{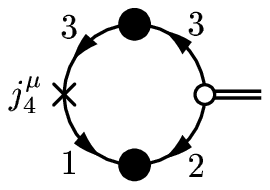}}
\end{center}
\caption{Fermion loop for the calculation of the type-II GB dispersion in the
three-flavor Fermi gas. The numbers adjacent to the fermion propagators
denote the respective flavors. The upper line represents the normal propagator of
the (ungapped) third fermion flavor, while the lower line represents the
pairing of the first two flavors.}
\label{fig:three_flavor_loop}
\end{figure}

To determine its dispersion relation, we need to evaluate the loop in Fig.
\ref{fig:three_flavor_loop} at nonzero external momentum $\vec k$. [In order to
make contact with literature \cite{He:2006ne}, we do not perform the Taylor
expansion in the external momentum as in Eq. \eqref{NR_dispersion_formula}.]
Note that since the type-II GB couples to two broken currents, $j^{\mu}_4$ and
$j^{\mu}_5$, we are free to choose any of them to calculate its dispersion.
Here, we take up $j^{\mu}_4$. The two propagators that appear in the loop are
given by the finite-temperature expressions (see Ref. \cite{He:2006ne})
$\mathcal G_{33} (i\nu_n,\vec k)=1/(i\nu_n-\xi_{\vec k})$, $\mathcal
F_{12}(i\nu_n,\vec k)=-i\Delta/[(i\nu_n)^2-E_{\vec k}^2]$, where $\Delta$ is
the gap parameter, $\xi_{\vec k}=\vec k^2/2m-\mu$ is the bare fermion energy
measured from the Fermi surface, and $E_{\vec k}=\sqrt{\xi_{\vec
k}^2+\Delta^2}$. The interpolating field for the GB is chosen as
$\psi^T\lambda_7\psi$. This may be understood as follows. By our choice, the
ground state condensate carries the third (anti)color of the representation
$\overline{\mathbf 3}$. The current $j^{\mu}_4$ whose transition amplitude we
calculate, then excites a difermion state with the first (anti)color. The GB
thus carries the quantum numbers of a pair of fermions with the second and the
third color, respectively. The same conclusion follows immediately from Eq.
\eqref{GB_fermion_coupling_Cooper}: The fermion self-energy is proportional to
$\lambda_2$ and $\{\lambda_2,\lambda_4\}=-\lambda_7$.

The point of using Eq. \eqref{loop_ratio} to calculate the GB dispersion
relation is that all common algebraic factors including the trace over the
flavor space cancel in the ratio. After stripping off such trivial factors, we
are thus left with the loop integral
\begin{equation}
J(\vec k)=\frac1\beta\sum_n\int\frac{d^3\vec
q}{(2\pi)^3}\doublet{1}{\tfrac{-\vec k+2\vec q}{2m}}\frac1{(i\nu_n)^2-E_{\vec
q-\vec k}^2}\frac1{i\nu_n-\xi_{\vec q}},
\label{loop_J}
\end{equation}
where the two options in the curly brackets stand for the insertion of the
temporal and spatial components of the current, respectively. Note that the
Matsubara frequency entering the loop graphs has been set to zero because, as
already remarked above, in case of a type-II GB the transition amplitude of
$j^0$ is nonzero even at vanishing energy.

After performing the Matsubara sum, Eq. \eqref{loop_J} acquires the form
\begin{equation}
J(\vec k)=\int\frac{d^3\vec q}{(2\pi)^3}\doublet{1}{\tfrac{-\vec k+2\vec
q}{2m}}\frac1{2E_{\vec q-\vec k}}
\left[\frac{1-f(E_{\vec q-\vec k})-f(\xi_{\vec
q})}{E_{\vec q-\vec k}+\xi_{\vec q}}+\frac{f(E_{\vec q-\vec k})-f(\xi_{\vec
q})}{E_{\vec q-\vec k}-\xi_{\vec q}}\right],
\label{three_flavor_result}
\end{equation}
where $f(x)=1/(e^{\beta x}+1)$ is the Fermi--Dirac distribution. This result is
completely equivalent to Eq. (A9) in Ref. \cite{He:2006ne}, though not
identical, of course, for we calculate a different loop graph here. The GB
dispersion may be directly inferred from here by means of Eq.
\eqref{loop_ratio}. We omit the explicit expression since it is rather
cumbersome and not particularly revealing.

\section{Four-flavor degenerate Fermi gas}
\label{Sec:four_flavor_gas}
In this section we are going to discuss the nonrelativistic
degenerate Fermi gas with four fermion flavors which, to our best knowledge,
has not been analyzed in literature so far. Besides being the next-to-simplest
nontrivial case of a nonrelativistic fermionic system exhibiting type-II GBs,
it also provides an interesting analogy with the two-color QCD with two quark
flavors \cite{Kogut:1999iv,Kogut:2000ek,Brauner:2006dv}.

We start by assuming the existence of four degenerate fermion species with a
common chemical potential $\mu$, which interact by a contact four-fermion
interaction invariant under the global $\mathrm{SU(4)\times U(1)}$ symmetry.
For the time being, all we need to know is that this interaction destabilizes
the Fermi sea by the formation of Cooper pairs. The detailed form of the
interaction will be specified later.

\subsection{Symmetry breaking patterns}
\label{Sec:symmetry_breaking_patterns} First we analyze the symmetry properties
of the ordered ground state. Analogously to the three-flavor case, the original
$\mathrm{SU(4)\times U(1)}$ symmetry is spontaneously broken by a difermion
condensate $\langle\psi_i\psi_j\rangle$, which transforms as an antisymmetric
rank-2 tensor of $\mathrm{SU(4)}$.

The analysis of the symmetry breaking patterns is much simplified by noting the
Lie algebra isomorphism $\mathrm{SU(4)\simeq SO(6)}$. The six-dimensional
antisymmetric tensor representation of $\mathrm{SU(4)}$ is equivalent to the
vector representation of $\mathrm{SO(6)}$. This correspondence is worked out in
detail in the Appendix. The order parameter is thus represented by a complex
$6$-vector, $\phi$, or equivalently, by two real $6$-vectors, $\phi_R=\re\phi$
and $\phi_I=\im\phi$. The situation is pretty much similar to that in spin-one
color superconductors where the order parameter is a complex $3$-vector of
$\mathrm{SO(3)}$ \cite{Buballa:2002wy,Schmitt:2003xq,Schmitt:2004et}.

There are two qualitatively different possibilities, distinguished by the
relative orientation of $\phi_R$ and $\phi_I$. First, if $\phi_R$ and $\phi_I$
are parallel, there is obviously an unbroken $\mathrm{SO(5)}$ rotational
invariance. The symmetry breaking pattern is $\mathrm{SO(6)\times U(1)\to
SO(5)}$, or $\mathrm{SU(4)\times U(1)\to Sp(4)}$, with six broken generators,
transforming as a $5$-plet and a singlet of the unbroken $\mathrm{SO(5)}$,
respectively. Note that in this case, the $\mathrm{U(1)}$ symmetry of the
Lagrangian may be used to even eliminate entirely the imaginary part of $\phi$,
i.e., to set $\phi_I=0$. Let us remark that it is this phase of the
nonrelativistic four-flavor Fermi gas that is analogous to the two-flavor
two-color QCD. There, the reality of $\phi$ is enforced by $C$ and $P$
conservation. Also, there is no global $\mathrm{U(1)}$ symmetry in two-color
QCD as the corresponding phase transformations are the axial ones, being broken
by the axial anomaly \cite{Kogut:1999iv,Kogut:2000ek}.

The more interesting, but also more complicated possibility is that with
nonzero $\phi_R$ and $\phi_I$ pointing in different directions. Then, the
$\mathrm{SO(6)\times U(1)}$ symmetry may be exploited to cast the order
parameter in the form $\phi=(\alpha,\beta+i\gamma,0,0,0,0)^T$, with real
$\alpha,\beta,\gamma$, and both $\alpha$ and $\gamma$ nonzero. In general the
$\mathrm{SO(6)\times U(1)}$ symmetry is now broken down to $\mathrm{SO(4)}$,
thus breaking $10$ generators, forming two vectors of the unbroken
$\mathrm{SO(4)}$ as well as two singlets. However, when $\beta=0$ and
$|\alpha|=|\gamma|$, there is an additional unbroken $\mathrm{U(1)}$ symmetry
generated by a combination of a rotation in the $(\phi_R,\phi_I)$ plane and the
original $\mathrm{U(1)}$ phase transformation. There is thus one less broken
generator and one less GB.

\subsection{Fermion propagator and excitation dispersions}
According to our general discussion above, we need to know the charge densities
in the ground state in order to determine the spectrum of GBs of the broken
symmetry. The charge densities, $\vev{\he\psi T_a\psi}$, are obtained from the
trace of the fermion propagator with the charge matrix $T_a$. At this place, we
have to specify the Lagrangian of the system.

Anticipating the Cooper pairing of the fermions, we choose the interaction
Lagrangian in a form suitable for a mean-field calculation,
\begin{equation}
\mathcal L=\he\psi\left(i\partial_t+\frac{\nabla^2}{2m}+\mu\right)\psi+G\sum_{i=1}^6|\psi^T\Sigma_i\psi|^2,
\label{Lagrangian}
\end{equation}
where $\psi$ is the four-component fermion field and $\Sigma_i$ are a set of
six basic antisymmetric $4\times4$ matrices, transforming as a vector of
$\mathrm{SO(6)}$, see Eqs. \eqref{matrix_orthogonality} and
\eqref{matrix_explicit}. Thus, the Lagrangian \eqref{Lagrangian} is manifestly
$\mathrm{SO(6)\simeq SU(4)}$ invariant.

Upon introduction of an auxiliary scalar field $\phi_i$ by the
Hubbard--Stratonovich transformation, the Lagrangian becomes
\begin{equation}
\mathcal L=\he\psi\left(i\partial_t+\frac{\nabla^2}{2m}+\mu\right)\psi-\frac1{4G}\he\phi_i\phi_i+
\frac12\psi^T\Sigma_i\psi\he\phi_i+\text{H.c.}.
\label{Lagrangian_HS}
\end{equation}
The vacuum expectation value of $\phi$ provides the order parameter for
symmetry breaking. In the mean-field approximation, $\phi$ is simply replaced
with its vacuum expectation value and the semi-bosonized Lagrangian
\eqref{Lagrangian_HS} then yields the inverse fermion propagator in the Nambu
space,
$$
G^{-1}=\left(
\begin{array}{cc}
(G_0^+)^{-1} & \Phi^-\\
\Phi^+ & (G_0^-)^{-1}
\end{array}\right),
$$
where $G_0^{\pm}$ are the bare propagators and $\Phi^-=\phi_i\he\Sigma_i$,
$\Phi^+=\he{(\Phi^-)}$. Inverting this block $2\times2$ matrix, we obtain
\cite{Schmitt:2004et}
$$
G=\left(
\begin{array}{cc}
G^+ & \Xi^-\\
\Xi^+ & G^-
\end{array}\right),
$$
where
\begin{equation}
G^{\pm}=\left[(G_0^{\pm})^{-1}-\Phi^{\mp}G_0^{\mp}\Phi^{\pm}\right]^{-1},\quad
\Xi^{\pm}=-G_0^{\mp}\Phi^{\pm}G^{\pm}.
\label{inverse_matrix_propagator}
\end{equation}

At finite temperature, the bare thermal propagators are simply
$G_0^+(i\nu_n,\vec k)=1/(i\nu_n-\xi_{\vec k})$ and $G_0^-(i\nu_n,\vec
k)=1/(i\nu_n+\xi_{\vec k})$. As explained in the previous section, we can
exploit the $\mathrm{SO(6)}$ symmetry of the Lagrangian \eqref{Lagrangian} so
that only two components of the $6$-vector $\phi$ are actually nonzero. Let us
denote them generally as $a,b$ ($a\neq b$), so that
$\Phi^-=\phi_a\he\Sigma_a+\phi_b\he\Sigma_b$. In the previous section we simply
took $a=1$ and $b=2$, but when later calculating the charge densities, it will
be more convenient to set $a=1$ and $b=6$.

According to Eq. \eqref{inverse_matrix_propagator}, the full mean-field fermion
propagator $G^-$ reads
\begin{equation}
G^-(i\nu_n,\vec k)=-(i\nu_n-\xi_{\vec k})(\nu_n^2+\xi_{\vec
k}^2+\Phi^+\Phi^-)^{-1}.
\label{Gminus_temp}
\end{equation}
Using the properties of the $\Sigma_i$ matrices we have
\begin{equation*}
\Phi^+\Phi^-=|\phi_a|^2+|\phi_b|^2-2i\im(\phi_a\phi_b^*)\Sigma_a\he\Sigma_b
=|\phi_a|^2+|\phi_b|^2-2\im(\phi_a\phi_b^*)[P_{ab}^{(+)}-P_{ab}^{(-)}].
\end{equation*}
The projectors $P^{(\pm)}_{ab}$ are defined in Eq.
\eqref{generator_projectors}. With projectors in the denominator, the matrix in
the expression \eqref{Gminus_temp} for $G^-$ can easily be inverted and yields
the final result
\begin{multline}
G^-(i\nu_n,\vec k)=-(i\nu_n-\xi_{\vec k})\frac{\nu_n^2+\xi_{\vec k}^2+|\phi_a|^2+|\phi_b|^2
+2\im(\phi_a\phi_b^*)[P_{ab}^{(+)}-P_{ab}^{(-)}]}{(\nu_n^2+\xi_{\vec k}^2+|\phi_a|^2+|\phi_b|^2)^2
-4[\im(\phi_a\phi_b^*)]^2}\\
=-(i\nu_n-\xi_{\vec k})\left[\frac{P_{ab}^{(+)}}{\nu_n^2+\xi_{\vec k}^2+|\phi_a|^2+|\phi_b|^2
-2\im(\phi_a\phi_b^*)}+\frac{P_{ab}^{(-)}}{\nu_n^2+\xi_{\vec k}^2+|\phi_a|^2+|\phi_b|^2
+2\im(\phi_a\phi_b^*)}\right].
\label{Gminus}
\end{multline}

From Eq. \eqref{Gminus} we can see that there are two doublets of fermionic
excitations over the ordered ground state, with generally nonzero gaps. Their
dispersion relations are most conveniently written in the form
\begin{equation}
\omega^{(\pm)}(\vec k)=\sqrt{\xi_{\vec k}^2+|\phi_a\mp i\phi_b|^2}.
\label{fermion_excitations}
\end{equation}
Two special cases deserve particular attention. First, in the $\mathrm{SO(5)}$
symmetric phase $\phi_b$ is set equal to zero so that all four excitations are
degenerate, forming a fundamental quadruplet of $\mathrm{SO(5)\simeq Sp(4)}$.
Second, in the $\mathrm{SO(4)\times U(1)}$ symmetric phase, $|\phi_a|=|\phi_b|$
with their complex phases differing by $\pi/2$. Eq. \eqref{fermion_excitations}
then implies that two of the fermions are actually gapless. Altogether we have
two pairs of degenerate fermions, which transform as doublets under the
respective $\mathrm{SU(2)}$ factors of $\mathrm{SO(4)\simeq SU(2)\times
SU(2)}$.

\subsection{Charge densities and GBs}
The ground state densities of the conserved charges are obtained from the
propagator $G^-$ as
\begin{equation*}
\vev{\he\psi T_a\psi}=-\frac1\beta\sum_n\int\frac{d^3\vec
k}{(2\pi)^3}\tr[T_a^TG^-(i\nu_n,\vec k)].
\end{equation*}
For sake of this section, it is convenient to fix the direction of the order
parameter $\phi$ in the $\mathrm{SO(6)}$ space. As already remarked above, we
choose $a=1$ and $b=6$. With the explicit form of the $\Sigma_i$ matrices given
in Eq. \eqref{matrix_explicit}, we find
\begin{equation*}
i\Sigma_1\he\Sigma_6=P_{16}^{(+)}-P_{16}^{(-)}=
-\left(\begin{array}{cc}
\tau_3 & 0\\
0 & \tau_3
\end{array}\right).
\end{equation*}
Since $i\Sigma_1\he\Sigma_6=M_{16}$ is one of the $\mathrm{SO(6)}$ generators
$M_{ij}$ defined in Eq. \eqref{M_definition}, the orthogonality relation
\eqref{M_orthogonality} implies that it can also be the only one which acquires
nonzero density in the ground state,
\begin{multline*}
\vev{M_{16}}=8\im(\phi_1\phi_6^*)\frac1\beta\sum_n\int\frac{d^3\vec
k}{(2\pi)^3}\frac{(i\nu_n-\xi_{\vec k})}
{(\nu_n^2+\xi_{\vec k}^2+|\phi_1|^2+|\phi_6|^2)^2
-4[\im(\phi_1\phi_6^*)]^2}\\
=-\int\frac{d^3\vec k}{(2\pi)^3}\xi_{\vec k}\left[ \frac{1-2f(\omega^{(+)}(\vec
k))}{\omega^{(+)}(\vec k)} -\frac{1-2f(\omega^{(-)}(\vec k))}{\omega^{(-)}(\vec
k)}\right],
\end{multline*}
after the Matsubara summation.

The spectrum of GBs may be understood as follows. First, in the
$\mathrm{SO(5)}$ symmetric phase, $\phi_6=0$ and there is no charge density in
the ground state. According to our general results, there are then six type-I
GBs---five corresponding to the coset $\mathrm{SO(6)/SO(5)}$, and one to the
broken $\mathrm{U(1)}$.

From now on, we shall analyze the $\mathrm{SO(4)}$ symmetric phase of the model
\eqref{Lagrangian}, i.e., assume that both $\phi_1$ and $\phi_6$ are nonzero
and have different complex phases. The condensate $\phi_1$ breaks all
$\mathrm{SO(6)}$ generators of the form $M_{1i}$, where $i=2,\dotsc,6$. On the
other hand, the condensate $\phi_6$ breaks all $\mathrm{SO(6)}$ generators of
the form $M_{j6}$, where $j=1,\dotsc,5$. This is altogether nine spontaneously
broken generators of $\mathrm{SO(6)}$. One of them, $M_{16}$, is a singlet of
the unbroken $\mathrm{SO(4)}$. The remaining eight ones group into two
$\mathrm{SO(4)}$ vectors, $M_{1i}$ and $M_{i6}$, with $i=2,\ldots,5$. Note
that, by Eq. \eqref{M_commutator}, $[M_{1i},M_{i6}]=2iM_{16}$ (no summation
over $i$!). Since $M_{16}$ has nonzero density in the ground state, the two
broken generators $M_{1i}$ and $M_{i6}$ couple to a type-II GB with a quadratic
dispersion relation. These eight broken generators are thus associated to an
$\mathrm{SO(4)}$ vector of type-II GBs.

Finally, for $\phi_6=\pm i\phi_1$ there is a residual unbroken $\mathrm{U(1)}$
symmetry generated by $\openone\pm M_{16}$. Note that the combinations
$M_{1i}\pm iM_{i6}$ are charged under this unbroken symmetry, carrying opposite
charges. It may also be illuminating to observe that, for instance, when
$\phi_6=-i\phi_1$, the matrix $\Phi^-$ has the form
$$
\Phi^-=2\phi_1\left(
\begin{array}{rrrr}
0 & 0 & 1 & 0\\
0 & 0 & 0 & 0\\
-1 & 0 & 0 & 0\\
0 & 0 & 0 & 0
\end{array}\right).
$$
In words, only the first and third fermion flavors participate in the pairing.
The second and fourth do not, and represent the two gapless excitations
predicted by Eq. \eqref{fermion_excitations}.

Let us now calculate the dispersion relation of the type-II GBs in this
$\mathrm{SO(4)\times U(1)}$ symmetric phase. (As will be shown in the next
section, the relation $\phi_6=\pm i\phi_1$ is one of the two possible solutions
of the mean-field gap equation, the other one corresponding to the
$\mathrm{SO(5)}$ symmetric phase.) As already noted above, there are four
type-II GBs, that couple to the broken generators $M_{1i}$ and $M_{i6}$. Since
these form a multiplet of the unbroken $\mathrm{SO(4)}$, they have all the same
dispersions. We choose $i=3$ for which the broken generators have a
particularly simple form,
$$
M_{13}=-\left(
\begin{array}{cc}
\tau_1 & 0\\
0 & \tau_1
\end{array}\right),\quad
M_{36}=\left(
\begin{array}{cc}
\tau_2 & 0\\
0 & \tau_2
\end{array}\right).
$$
Together with $M_{16}$ these generate an $\mathrm{SU(2)}$ subalgebra of
$\mathrm{SU(4)}$.

\begin{figure}
\begin{center}
\scalebox{1}{\includegraphics{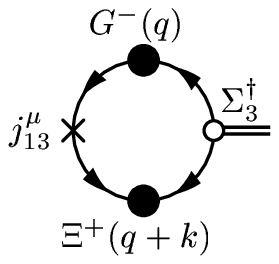}}
\end{center}
\caption{Fermion loop for the calculation of the type-II GB dispersion in the
four-flavor Fermi gas described by the Lagrangian \eqref{Lagrangian}.}
\label{fig:four_flavor_loop}
\end{figure}

The generators $M_{13}$ and $M_{36}$ must both couple to the sought type-II GB.
When acting on the ground state, they both excite the third component of
$\phi$, so the interpolating field for the GB will be $\psi^T\Sigma_3\psi$. In
order to determine the GB dispersion, we evaluate its coupling to the broken
current of $M_{13}$, which is given by the loop graph in Fig.
\ref{fig:four_flavor_loop}, analogous to that for the thee-flavor case in Fig.
\ref{fig:three_flavor_loop}.
After performing the Matsubara sum, and throwing away trivial numerical
factors, we arrive at the loop integral, analogous to Eq.
\eqref{three_flavor_result},
\begin{equation*}
J(\vec k)=\int\frac{d^3\vec q}{(2\pi)^3}\doublet{1}{\tfrac{-\vec k+2\vec
q}{2m}}\frac{\xi_{\vec q}}{2\omega^{(\pm)}_{\vec q-\vec k}\omega^{(\mp)}_{\vec q}}
\left[\frac{1-f(\omega^{(\pm)}_{\vec q-\vec k})-f(\omega^{(\mp)}_{\vec
q})}{\omega^{(\pm)}_{\vec q-\vec k}+\omega^{(\mp)}_{\vec
q}}+\frac{f(\omega^{(\pm)}_{\vec q-\vec k})-f(\omega^{(\mp)}_{\vec
q})}{\omega^{(\pm)}_{\vec q-\vec k}-\omega^{(\mp)}_{\vec q}}\right],
\end{equation*}
where the upper and lower signs were obtained with $\phi_6=\pm i\phi_1$,
respectively. Ultimately, both yield the same dispersion, which is found by the
Taylor expansion to first order in momentum and plugging the result into Eq.
\eqref{quadratic_dispersion}.

\subsection{Gap equation}
The actual value of the order parameter $\phi$ is determined by the gap
equation. To that end, we need to know the anomalous part of the matrix
propagator which follows from Eq. \eqref{inverse_matrix_propagator},
$$
\Xi^-(i\nu_n,\vec k)=-\frac1{i\nu_n-\xi_{\vec
k}}(\phi_a\he\Sigma_a+\phi_b\he\Sigma_b)G^-(i\nu_n,\vec k).
$$
In the mean-field approximation, we require the self-consistency at the
one-loop level, which leads to the gap equation
\begin{equation*}
\Phi^-=2G\frac1\beta\sum_n\int\frac{d^3\vec k}{(2\pi)^3}\he\Sigma_i
\tr[\Sigma_i\Xi^-(i\nu_n,\vec k)].
\end{equation*}
After working out the algebraic trace and the Matsubara sum, this set of gap
equations for $\phi_a$ and $\phi_b$ may be rewritten in a particularly
symmetric form,
\begin{equation}
\begin{split}
\phi_a+i\phi_b&=4G(\phi_a+i\phi_b)I^{(-)}(\phi),\\
\phi_a-i\phi_b&=4G(\phi_a-i\phi_b)I^{(+)}(\phi),
\end{split}
\label{AB_symmetric_eqs}
\end{equation}
where
$$
I^{(\pm)}(\phi)=\int\frac{d^3\vec k}{(2\pi)^3}\frac {1-2f(\omega^{(\pm)}(\vec
k))}{\omega^{(\pm)}(\vec k)}.
$$

The gap equations \eqref{AB_symmetric_eqs} apparently admit two qualitatively
different solutions. The first one corresponds to $\im(\phi_a\phi_b^*)=0$,
i.e., the $\mathrm{SO(5)}$ symmetric phase. The gap equation then simply reads
\begin{equation}
1=4GI^{(\pm)}(\phi).
\label{gapeq_A}
\end{equation}
Either sign can be used in the integrals since, in this case, the
$\omega^{(\pm)}$ dispersions are degenerate.

The second solution has $\im(\phi_a\phi_b^*)\neq0$. Eq.
\eqref{AB_symmetric_eqs} then implies that $\phi_a=\pm i\phi_b$. For the two
possibilities, the gap equations are
\begin{equation}
\begin{split}
1&=4GI^{(-)}(\phi),\quad\text{for $\phi_a=+i\phi_b$,}\\
1&=4GI^{(+)}(\phi),\quad\text{for $\phi_a=-i\phi_b$.}
\end{split}
\label{gapeq_B}
\end{equation}

Note that this solution of the gap equations \eqref{AB_symmetric_eqs} precisely
corresponds to the $\mathrm{SO(4)\times U(1)}$ phase discussed in Sec.
\ref{Sec:symmetry_breaking_patterns}. In either case, $\phi_a=\pm i\phi_b$,
only the two gapped fermion flavors contribute to the gap equation
\eqref{gapeq_B}.

It is interesting to observe that the mean-field analysis does not support the
most general form of the order parameter $\phi$. Rather, either the real and
imaginary parts of $\phi$ are parallel, breaking the $\mathrm{SO(6)\times
U(1)}$ symmetry to $\mathrm{SO(5)}$, or they are perpendicular with the same
magnitude, breaking the symmetry to $\mathrm{SO(4)\times U(1)}$.

Of course, which of these two possibilities is the actual ground state may
depend on the magnitude of the coupling constant and the chemical potential,
and can only be decided by solving the gap equations \eqref{gapeq_A} and
\eqref{gapeq_B} and comparing their free energies. This task is however, beyond
the scope of the present paper where the model \eqref{Lagrangian} only serves
as an illustration of the general features of symmetry breaking in presence of
charge density.

\section{Pseudo-Goldstone bosons}
\label{Sec:pseudo_GBs}
So far we have been dealing exclusively with exact symmetries for which, once
spontaneously broken, the Goldstone theorem ensures the existence of exactly
gapless excitations. Real world is, however, more complicated. Almost as a
rule, global internal symmetries are usually only approximate, leading to
approximately conserved currents and, when spontaneously broken, to the
existence of excitations with a small gap, so called pseudo-Goldstone bosons
(pGBs). (In the following, we shall keep using the term Goldstone boson unless
the existence of a nonzero gap is important for our discussion.)

Let us just note that, in fact, the only generic type of an exact global
symmetry is the particle number conservation. This is, however, only Abelian
and hence does not lead to the nontrivial phenomena studied in this paper. In
addition, gauge symmetries are always exact. Here, once the gauge is
appropriately fixed and the remaining global symmetry spontaneously broken,
there are no physical GB states as they are ``eaten'' by the gauge bosons,
making them massive by the Higgs mechanism. It turns out that the number of
massive gauge bosons is always equal to the number of broken generators even if
the number of would-be GBs is smaller \cite{Gusynin:2003yu}.

\subsection{Counting of pseudo-Goldstone bosons}
It should be stressed that for approximate symmetries, there is no rigorous
theorem analogous to the Goldstone's, which would guarantee the existence of
``approximately gapless'' excitations. On a heuristic level, however, the
conclusions of the theorems of Goldstone and of Nielsen and Chadha may be
carried over to this less perfect, yet more common and physical situation.
Before a more detailed discussion of approximate symmetries, let us illustrate
this on a simple example.

Both Goldstone and Nielsen--Chadha theorems make predictions about the
low-momentum limit of the GB dispersion relation without actually specifying
what the low momentum means. What is the type and number of pGBs associated
with a spontaneously broken approximate symmetry, therefore depends on the
scale at which the symmetry is broken explicitly. Consider the simple model
studied in Sec. \ref{Sec:Miransky_model}. The type-II GB has a massive partner
which is annihilated by $\he\phi_1$, i.e., its antiparticle, and their exact
tree-level dispersions may be jointly expressed as
$$
\omega=\sqrt{\vec k^2+\mu^2}\pm\mu,
$$
see Eq. \eqref{typeIIdispersion}. At low momentum, we clearly see a gapless
mode with a quadratic dispersion and a mode with the gap $2\mu$. On the other
hand, at high momentum both modes have almost linear dispersion.

So, once the $\mathrm{SU(2)\times U(1)}$ symmetry of the model
\eqref{Lagrangian_Shovkovy} is explicitly broken, the resulting pGB spectrum
depends on whether the scale of the explicit breaking is smaller or larger than
the chemical potential $\mu$. For weak breaking, we still see an approximately
gapless mode with a quadratic dispersion and a mode with a gap approximately
$2\mu$. On the other hand, for strong explicit breaking, larger than $\mu$, we
can no longer resolve the quadratic onset of the dispersion, and instead find
two pGBs with linear dispersion relations. Note that in both cases the
Nielsen--Chadha counting rule is satisfied even for pGBs.

\subsection{Ward identity and dispersion relations}
Let us now consider a symmetry which is explicitly broken by adding a small
term to the Lagrangian of the theory. The associated current then becomes only
approximately conserved and we find $\partial_{\mu}j_a^{\mu}=\Delta_a\mathcal L$, where
$\Delta_a\mathcal L$ is the variation of the Lagrangian under the symmetry
transformation. The basic assumption of an ``approximate'' spontaneously broken
symmetry is that there is still a mode $\ket{\pi(\vec k)}$ which couples to the
current $j_a^{\mu}$. The parameterization
\eqref{current_amplitude_parametrization}, which is based just on rotational
invariance, thus still holds. What does change is, of course, Eq.
\eqref{general_dispersion} which now becomes
\begin{equation}
(\omega^2-\vec k^2)F_a+\omega G_a=\bra0\Delta_a\mathcal L\ket{\pi(\vec k)}.
\label{general_dispersion_explicit}
\end{equation}

Up to this change, we proceed along the same line of argument as in Sec.
\ref{Sec:general}. The Ward identity \eqref{Ward_identity} acquires a new term
that accounts for the explicit symmetry breaking,
\begin{equation*}
\partial^x_{\mu}\bra0T\{j_a^{\mu}(x)j_b^{\nu}(y)\}\ket0=\bra0T\{\Delta_a\mathcal L(x)j^{\nu}_b(y)\}\ket0
+\delta^4(x-y)if_{abc} \bra0j^{\nu}_c(x)\ket0.
\end{equation*}
Since both the broken currents and $\Delta_a\mathcal L$ couple to the GB state,
both correlators can be expressed via the K\"all\'en--Lehmann representation.
We find that cancelation of the one-particle poles exactly requires the energy
and momentum to fulfill the dispersion relation
\eqref{general_dispersion_explicit}. In the limit $|\vec k|\to0$, the Ward
identity leads to the modified density rule,
$$
\im[G_aG_b^*+\omega(0)G_aF_b^*]=\frac12f_{abc}\bra0j_c^0\ket0,
$$
cf. Eq. \eqref{density_rule}. The explicit breaking of the symmetry enters here
in terms of the (nonzero) energy gap, $\omega(0)$.

Eq. \eqref{general_dispersion_explicit} may be used to give a precise meaning
to the otherwise vague term ``approximately gapless mode'', as well as to
distinguish type-I and type-II GBs in the case of an explicitly broken
symmetry. Let us thus, again, assume that $G_a(|\vec k|)$ has a nonzero limit
as $|\vec k|\to0$, and denote $\bra0\Delta_a\mathcal L\ket{\pi(\vec k)}=H_a(|\vec k|)$,
for sake of brevity.

According to Eq. \eqref{general_dispersion_explicit}, the energy gap
$\omega(0)$ of the pGB is given by the solution of the equation
$$
\omega^2(0)F_a+\omega(0)G_a=H_a,
$$
which has exactly one positive solution provided all coefficients,
$F_a,G_a,H_a$, are positive,
$$
\omega(0)=\frac1{2F_a}\left(-G_a+\sqrt{G_a^2+4F_aH_a}\right).
$$

The low-momentum behavior of the GB dispersion is apparently determined by the
relative value of $G_a^2$ and $F_aH_a$. If the explicit symmetry breaking is
weak enough, i.e., if $F_aH_a\ll G_a^2$, the term quadratic in $\omega$ may be
neglected in comparison with that linear in $\omega$, and the low-momentum
dispersion relation of the pGB reads
\begin{equation}
\omega(\vec k)=\frac{H_a}{G_a}+\vec k^2\frac{F_a}{G_a},
\label{quadratic_dispersion_explicit}
\end{equation}
as opposed to Eq. \eqref{quadratic_dispersion}. We thus find a type-II pGB.

If, on the other hand, $F_aH_a\gg G_a^2$, the $\omega^2$ term dominates over
the $\omega$ term in Eq. \eqref{general_dispersion_explicit}, and the resulting
low-momentum dispersion relation,
\begin{equation}
\omega(\vec k)=\sqrt{\frac{H_a}{F_a}+\vec k^2},
\label{linear_dispersion_explicit}
\end{equation}
is one of a type-I pGB with nonzero mass squared $H_a/F_a$.

\subsection{Nambu--Jona-Lasinio model}
The formulas
\eqref{general_dispersion_explicit}--\eqref{linear_dispersion_explicit} are
absolutely general but, unfortunately, not of much use unless we are somehow
able to calculate the new amplitude, $\bra0\Delta_a\mathcal L\ket{\pi(\vec k)}$.
Nevertheless, there is a wide class of systems where this can be determined in
the same way as the current amplitudes $F_a$ and $G_a$: They are the models of
the NJL type where the symmetry-breaking operator is bilinear in the fermion
fields.

First of all, note that this is all we need, for instance, to provide a
realistic description of dense quark matter since both quark masses and the
various chemical potentials can be taken into account by adding appropriate
bilinear operators into the Lagrangian.

The point is that in such a case, the dispersion relation of the GB
\eqref{quadratic_dispersion_explicit} [or \eqref{linear_dispersion_explicit} as
well] is given by the ratio of fermion loops, just like in Eq.
\eqref{loop_ratio}, which now becomes
\begin{equation*}
\omega(\vec k)=\frac{\Delta_a\mathcal L\scalebox{0.8}{\parbox{20mm}{\includegraphics{fig1}}}\:\:+
\vec k\cdot\vec j_a\scalebox{0.8}{\parbox{20mm}{\includegraphics{fig1}}}}
{j^0_a\scalebox{0.8}{\parbox{20mm}{\includegraphics{fig1}}}}.
\end{equation*}
All that was said in Sec. \ref{Sec:fermion_loops} then holds without change
also in the presence of explicit symmetry breaking.

\section{Conclusions}
In this paper, we analyzed the general problem of spontaneous symmetry breaking
in presence of nonzero charge density. Based on a Ward identity for the broken
symmetry, we showed in Sec. \ref{Sec:general} that nonzero density of a
commutator of two broken currents implies the existence of a GB with nonlinear
dispersion relation. Under fairly general circumstances, this dispersion is
quadratic. The only assumption is the existence of a nonzero low-momentum limit
of the transition amplitude of the spatial current, $F_a$.

It is interesting to compare this result to that of Leutwyler
\cite{Leutwyler:1994gf}. In his approach, nonzero density of a non-Abelian
charge leads to the presence of a term with a single time derivative in the
effective Lagrangian. Together with the standard two-derivative kinetic term,
this results in a quadratic dispersion of the GB. In fact, however, Leutwyler's
kinetic term is proportional to the same constant $F_a$ that we use here just
to parameterize the current transition amplitude. Thus, regarding the
possibility of the existence of GBs with energy proportional to some higher
power of momentum, our results are equivalent. Nevertheless, the comparison
with Leutwyler's approach gives us new insight into this open question: In
order to have a GB with the power of momentum higher than two in the dispersion
relation, the standard kinetic term in the low-energy effective Lagrangian
would have to vanish. It is hard to imagine how such a theory would be defined
even perturbatively, yet such a possibility is not theoretically ruled out.

In Sec. \ref{Sec:linear_sigma_model} we demonstrated the general results of
Sec. \ref{Sec:general} on the linear sigma model, which we studied in detail in
our previous work \cite{Brauner:2005di,Brauner:2006xm}. We were thus able to
prove on a very general ground, that the tree-level dispersion of type-II GBs
in such a class of models is of the form \eqref{typeII_dispersion_Q}.

While in the framework of the linear sigma model we merely verified results
that could be obtained in a different manner, in the models of the NJL type our
general results are practically useful. In Sec. \ref{Sec:NJL_model} we
suggested to calculate the GB dispersion relation by a direct evaluation of its
coupling to the broken current. This procedure considerably reduces both the
length and the complexity of the analytic computation as compared with the
conventional approach using the Hubbard--Stratonovich transformation and the
consecutive diagonalization of the quadratic part of the action.

In Sec. \ref{Sec:four_flavor_gas} we analyzed a particular example of a system
exhibiting type-II GBs: The nonrelativistic gas of four fermion flavors with an
$\mathrm{SU(4)\times U(1)}$ global symmetry. We showed that the mean-field gap
equation admits two qualitatively different solutions, i.e., two different
ordered phases. In the first one, all fermion flavors pair in a symmetric way,
leaving unbroken an $\mathrm{SO(5)}$ subgroup of the original symmetry. The six
broken generators give rise to six type-I GBs.

In the other phase, only two of the four fermion flavors participate in the
pairing. The other two flavors remain gapless. In this case, the unbroken
symmetry is $\mathrm{SO(4)\times U(1)}$ and one of the broken
charges---corresponding to the difference of the number of the paired and
unpaired fermions---acquires nonzero density. Consequently, eight of the nine
broken generators give rise to four type-II GBs with a quadratic dispersion.
Which of these two phases is the actual ground state of the system, can only be
decided upon the solution of the gap equation.

Finally, in Sec. \ref{Sec:pseudo_GBs} we extended the results achieved in the
preceding parts of the paper to spontaneously broken approximate symmetries. We
discovered that once the notions of the type and low-momentum dispersion
relation of the GB are defined and treated carefully, the general results
remain valid. In particular, in the NJL model this allows their application to
the realistic description of dense quark matter.

\begin{acknowledgments}
The author is indebted to M. Buballa and J. Ho\v{s}ek for fruitful discussions
and critical reading of the manuscript. He is also obliged to J. Noronha, D.
Rischke, and A. Smetana for discussions. Furthermore, he gratefully
acknowledges the hospitality offered him by ECT$^*$ Trento and by the Frankfurt
Institute for Advanced Studies, where parts of the work were performed. The
present work was in part supported by the Institutional Research Plan
AV0Z10480505 and by the GA CR grant No. 202/06/0734.
\end{acknowledgments}

\appendix
\section{Algebra of antisymmetric $4\times4$ matrices}
In Ref. \cite{Brauner:2006dv} we showed that a general unitary antisymmetric
$4\times4$ matrix $\Sigma$ with unit determinant may be cast in the form
$\Sigma=n_i\Sigma_i$, $i=1,\dotsc,6$, where $\Sigma_i$ is a set of six basis matrices,
satisfying the constraint
\begin{equation}
\he\Sigma_i\Sigma_j+\he\Sigma_j\Sigma_i=2\delta_{ij}\openone,
\label{matrix_orthogonality}
\end{equation}
and $n_i$ is an either real or pure imaginary unit vector. By this construction we
provided an explicit mapping between the cosets $\mathrm{SU(4)/Sp(4)}$ and
$\mathrm{SO(6)/SO(5)}$. We also found a particular explicit solution to Eq.
\eqref{matrix_orthogonality},
\begin{equation}
\begin{split}
\Sigma_1=\left(
\begin{array}{cc}
0 & -1\\
1 & 0
\end{array}\right),\quad
\Sigma_2=\left(
\begin{array}{cc}
\tau_2 & 0\\
0 & \tau_2
\end{array}\right),\quad
\Sigma_3=\left(
\begin{array}{cc}
0 & i\tau_1\\
-i\tau_1 & 0
\end{array}\right),\\
\Sigma_4=\left(
\begin{array}{cc}
i\tau_2 & 0\\
0 & -i\tau_2
\end{array}\right),\quad
\Sigma_5=\left(
\begin{array}{cc}
0 & i\tau_2\\
i\tau_2 & 0
\end{array}\right),\quad
\Sigma_6=\left(
\begin{array}{cc}
0 & i\tau_3\\
-i\tau_3 & 0
\end{array}\right).
\end{split}
\label{matrix_explicit}
\end{equation}

By a simple complex conjugation, using the antisymmetry of the $\Sigma_i$s, Eq.
\eqref{matrix_orthogonality} translates to
\begin{equation}
\Sigma_i\he\Sigma_j+\Sigma_j\he\Sigma_i=2\delta_{ij}\openone.
\label{matrix_orthogonality_2}
\end{equation}

In the context of the present paper, $n_i$ no is longer a unit vector, but
rather an arbitrary complex six-dimensional vector. We start by showing that
its real and imaginary parts realize independent real representations of the
group $\mathrm{SU(4)\simeq SO(6)}$. This acts on $\Sigma$ as $\Sigma\to U\Sigma
U^T$. For an infinitesimal transformation, $U=\openone+iX$ with a traceless,
Hermitian $X$. Using the orthogonality relation
\begin{equation}
\tr\he\Sigma_i\Sigma_j=4\delta_{ij}, \label{matrix_orthogonality_trace}
\end{equation}
we derive the change of the coordinate $n_i$ induced by the transformation $U$,
$$
\delta n_i=\frac{in_j}4\tr[\he\Sigma_i(X\Sigma_j+\Sigma_jX^T)]=
\frac{in_j}2\tr(X\Sigma_j\he\Sigma_i).
$$
(We transposed the second term and used the antisymmetry of $\Sigma_i$.) Now
for traceless $X$, the matrix $i\tr(X\Sigma_j\he\Sigma_i)$ is, with a little
help from Eq. \eqref{matrix_orthogonality_2}, readily seen to be real. As an
immediate consequence, the real part of $n_i$ transforms into the real part and
analogously for the imaginary part, as was to be proven.

In fact, just the six matrices $\Sigma_i$ generate the whole algebra of $\mathrm{SU(4)}$.
To see this, note that $i\Sigma_i\he\Sigma_j$, $i\neq j$,
are $15$ independent matrices with
the following properties: 1. They are Hermitian. 2. Their trace is zero.
3. Their square is equal to unit matrix.

Proof. Hermiticity follows from
$\he{(i\Sigma_i\he\Sigma_j)}=-i\Sigma_j\he\Sigma_i=i\Sigma_i\he\Sigma_j$, by
Eq. \eqref{matrix_orthogonality_2}. The same identity together with the
unitarity of $\Sigma_i$ implies
$(i\Sigma_i\he\Sigma_j)^2=-\Sigma_i\he\Sigma_j\Sigma_i\he\Sigma_j=
\Sigma_i\he\Sigma_j\Sigma_j\he\Sigma_i=\openone$.
The tracelessness is an immediate
consequence of Eq. \eqref{matrix_orthogonality_trace}.

The fact that the matrices $i\Sigma_i\he\Sigma_j$ square to one tells us that
their eigenvalues are $\pm1$ and, moreover, both with the twofold degeneracy, in
order to ensure zero trace. We may then write
$i\Sigma_i\he\Sigma_j=P_{ij}^{(+)}-P_{ij}^{(-)}$, where the projectors
$P_{ij}^{(\pm)}$ are conveniently expressed as
\begin{equation}
\begin{split}
P_{ij}^{(+)}&=(\Sigma_i-i\Sigma_j)\he{(\Sigma_i-i\Sigma_j)},\\
P_{ij}^{(-)}&=(\Sigma_i+i\Sigma_j)\he{(\Sigma_i+i\Sigma_j)}.
\end{split}
\label{generator_projectors}
\end{equation}

The identity \eqref{matrix_orthogonality_2} implies that the product
$i\Sigma_i\he\Sigma_j$ is, for $i\neq j$, antisymmetric under the exchange of
$i$ and $j$. To make this antisymmetry explicit, let us define a new set of
matrices,
\begin{equation}
M_{ij}=\frac
i2(\Sigma_i\he\Sigma_j-\Sigma_j\he\Sigma_i)=i(\Sigma_i\he\Sigma_j-\delta_{ij}\openone).
\label{M_definition}
\end{equation}
By means of Eqs. \eqref{matrix_orthogonality} and
\eqref{matrix_orthogonality_2}, these may be shown to satisfy the commutation
relation
\begin{equation}
[M_{ij},M_{kl}]=2i(\delta_{il}M_{jk}+\delta_{jk}M_{il}-\delta_{ik}M_{jl}-\delta_{jl}M_{ik}).
\label{M_commutator}
\end{equation}
The matrices $M_{ij}$ thus provide an explicit mapping between the Lie algebras
of $\mathrm{SU(4)}$ and $\mathrm{SO(6)}$.

Finally, Eqs. \eqref{matrix_orthogonality}, \eqref{matrix_orthogonality_2}, and
\eqref{matrix_orthogonality_trace} imply the identity, analogous to that for
the Dirac $\gamma$ matrices,
\begin{equation*}
\tr(\Sigma_i\he\Sigma_j\Sigma_k\he\Sigma_l)=4(\delta_{ij}\delta_{kl}-\delta_{ik}\delta_{jl}+
\delta_{il}\delta_{jk}).
\end{equation*}
As a consequence we have the orthogonality relation for the $\mathrm{SO(6)}$ generators
$M_{ij}$,
\begin{equation}
\tr(M_{ij}M_{kl})=4(\delta_{ik}\delta_{jl}-\delta_{il}\delta_{jk}).
\label{M_orthogonality}
\end{equation}

Note also that none of the identities proven in this Appendix relies on the
explicit representation of the matrices $\Sigma_i$, Eq.
\eqref{matrix_explicit}. All the results thus hold independently of the
particular realization of $\Sigma_i$. In a sense, the construction worked out
here is similar to that of the Clifford algebra of the spinor representations
of the orthogonal groups. However, in case of $\mathrm{SO}(2n)$, this is
realized by matrices of rank $2^n$ \cite{Georgi:1982jb}, while here we need
just $4\times4$ matrices for the group $\mathrm{SO(6)}$.

\end{document}